\documentclass{aastex}

\shorttitle{S and Ne in Halo Planetary Nebulae}
\shortauthors{Dinerstein et al.}

\begin{document}

\title{Observations of [S IV] 10.5 $\mu$m and [Ne II] 12.8 $\mu$m
in Two Halo Planetary Nebulae: Implications for Chemical Self-Enrichment} 

\author{Harriet L. Dinerstein\altaffilmark{1,2}, 
Matthew J. Richter\altaffilmark{1,2,3}, John H. Lacy\altaffilmark{1,2}, 
and K. Sellgren\altaffilmark{4}}

\altaffiltext{1}{Astronomy Department, The University of Texas at Austin,
  1 University Station C1400, Austin, TX 78712-0259;  
  harriet@astro.as.utexas.edu,
  lacy@astro.as.utexas.edu}

\altaffiltext{2}{Visiting Astronomer at the Infrared Telescope Facility, which is
  operated by the University of Hawaii under contract with the National 
  Aeronautics and Space Administration.}

\altaffiltext{3}{now at: Physics Department, 1 Shields Avenue,
  University of California, Davis, CA 95616; 
  richter@physics.ucdavis.edu}

\altaffiltext{4}{Department of Astronomy, Ohio State University, 
  140 W. 18th Ave., Columbus, OH 43210; sellgren@astronomy.ohio-state.edu}

\begin{abstract}

We have detected the [\ion{S}{4}] 10.5 $\mu$m and
[\ion{Ne}{2}] 12.8 $\mu$m fine-structure lines
in the halo population planetary nebula (PN)
DdDm~1, and set upper limits on their intensities
in the halo PN H~4-1. We also present new 
measurements of optical lines from various
ions of S, Ne, O, and H for DdDm~1, based on
a high-dispersion spectrum covering the spectral
range 3800 \AA\ -- 1 $\mu$m.
These nebulae have similar O/H abundances,
(O/H) $\sim$ 1 $\times$ 10$^{-4}$, but S/H and Ne/H 
are about half an order of magnitude lower in H~4-1 than in
DdDm~1; thus H 4-1 appears to belong to a more
metal-poor population. 
This supports previous suggestions
that PNe arising from  
metal-poor progenitor stars can have elevated oxygen
abundances due to internal nucleosynthesis
and convective dredge-up. 
It is generally accepted that high abundances
of carbon in many PNe results from self-enrichment.
To the extent that oxygen can also be affected, 
the use of nebular O/H values to infer the overall
metallicity of a parent stellar population 
(for example, in external galaxies) may
be suspect, particularly for low metallicities.

\end{abstract}

\keywords{infrared: lines---planetary nebulae: general---planetary nebulae:
individual (DdDm~1, H~4-1)---nucleosynthesis, abundances---stars: AGB and post-AGB}

\section{INTRODUCTION}

Of the $\ge$ 1200 catalogued planetary nebulae (PNe) in our Galaxy, less than a
dozen are truly metal-poor. These halo or ``Type~IV'' PNe, in the terminology of 
Peimbert (1978, 1990), have (O/H) abundances more than a factor of 4 below solar, 
even allowing for the notorious disparity between stellar and nebular 
abundance scales due to uncertainties in nebular electron temperature structure
(e.g. Peimbert 1967, 1995; Dinerstein 1990). 
Not surprisingly, this small sample has
been the target of intense scrutiny 
in order to determine their chemical compositions.
These studies have revealed that the ratios among the heavy elements in the halo PNe  
are non-solar and differ from object to object
(e.g. Torres-Peimbert, Rayo, \& Peimbert 1981; Pe\~{n}a et al. 1992; 
Howard, Henry, \& McCartney 1997). 
The heavier $\alpha$-capture elements such as Ar and S are more
deficient than the CNO group, and it has been
argued by a number of authors that the former are more indicative
of the initial stellar composition
(Barker 1980, 1983; Torres-Peimbert et al. 1981).   
However, since some important ions do not give rise
to optical or UV emission lines, having ground configurations
that produce only infrared fine-structure lines, 
the total abundances of these elements may be under- or over-estimated 
when these ions are not directly observed (e.g. Dinerstein 1995).
This circumstance has motivated a number of studies of infrared emission lines
in PNe using ground-based and space observatories
(Dinerstein 1980a, 1980b; Beck
et al. 1981; Pottasch et al. 1986; Liu et al. 2001). 

Garnett \& Lacy (1993) attempted to measure the [\ion{S}{4}] 10.5 $\micron$
line in two of the most metal-poor halo PNe, Ps~1 (also known as K648) and 
BB~1 (BoBn~1), 
to test the possibility that
abundances inferred from the optical lines alone, which sample only [\ion{S}{2}]
and [\ion{S}{3}], might be too low. Using Irshell (Lacy et al. 1989), a fore-runner
of the instrument we used here, they set stringent upper limits on the [\ion{S}{4}]
line fluxes and confirmed that the S/H abundances in these objects are indeed
much lower than O/H. This result was consistent with earlier determinations
of S/H by Barker (1983) and others, and with the suggestion that elements such
as Ar and S provide a more reliable benchmark of the composition of 
the progenitor star, whereas O/H and possibly Ne/H may be enhanced by
internal processing  
(Torres-Peimbert \& Peimbert 1979; Barker 1980, 1983; Dopita et al. 1997).   

In this paper, we present measurements of [\ion{S}{4}] 10.5 $\micron$ 
and [\ion{Ne}{2}] 12.8 $\micron$ in two
additional halo PNe, DdDm~1 and H~4-1, in addition to new optical
line measurements of the former. 
Our detections of the infrared lines in DdDm~1 and upper limits for their
fluxes in H~4-1 confirms the idea that there are two classes 
of PNe with the same range of O/H values: modestly metal-poor objects
with normal heavy-element ratios, and extremely metal-poor objects in which 
O/H has been self-enriched during the AGB phase.

\section{OBSERVATIONS AND REDUCTIONS}

\subsection{Mid-IR Lines}

The mid-IR observations were obtained with TEXES, the Texas
Echelon Cross Echelle Spectrograph (Lacy et al. 2002),
at NASA's Infrared Telescope Facility (IRTF)
on Mauna Kea, in late June 2001.  
A 3.0$^{\prime \prime}$-wide slit was employed
with slit lengths of $\sim$ 7.5$^{\prime \prime}$ and 
9$^{\prime \prime}$ respectively
at 10 and 12 $\mu$m. 
These apertures are large enough to contain essentially
all of the line emission from these compact nebulae.
The weather conditions were only fair,
but the observed lines were Doppler shifted to relatively clean
parts of the atmosphere; the radial velocity of DdDm~1 is
$\sim$ --304~km~s$^{-1}$ (Barker \& Cudworth 1984; herefter B84) and
that of H~4-1 is $\sim$ --141~km~s$^{-1}$ (Miller 1969).
All of the observations were obtained at air masses 
less than 1.2. 
We nodded the telescope 4$^{\prime \prime}$ or 
5$^{\prime\prime}$ along the slit in
order to increase the observing efficiency, with dwell times
of 3-12 seconds per beam position.  Total integration times are
given in Table~1.
The objects were identified visually using the telescope on-axis camera, and
offset guiding was used while observing.

The data were reduced using the standard TEXES pipeline (Lacy et al. 2002).
We removed the background emission by
taking differences between target frames and blank sky frames, and
corrected for pixel response variations and atmospheric effects by
subtracting a sky frame
from a blackbody frame to create a flat field, first-order atmospheric
correction, and flux calibration.  These frames are taken every 3 to 10 minutes.
The zero point for the wavelength scale of each observation set was determined 
by specifying the pixel value and wavelength
of one atmospheric line in one order.
This process typically is accurate to better than 1~km/s across the entire
observed region.
After flatfielding, we resampled the array so that the spectral and
spatial dimensions are orthogonal and run along rows and columns.
We extracted the final spectra by summing over 
$\sim$2.5$\arcsec$ on each side of the nod. 

Flux calibrations were performed using similarly extracted spectra of
$\alpha$~Boo and $\delta$~Vir.  
Where no line was detected, we set upper limits (3$\sigma$)
by summing over spectral regions corresponding to 
the width of the detected lines in DdDm~1.
For an object which uniformly fills the aperture, the
resolving power of the 3.0$\arcsec$ slit is $\sim$ 37,000; for
DdDm~1, which is nearly a point source (B84), the effective
resolution can be roughly twice as high, under good seeing conditions.
The spectra for DdDm~1 are shown in Fig. 1; the ionic lines are
strong enough that they were clearly visible in each nod pair.
We also attempted to measure \ion{H}{1} 7--6 (Hu~$\alpha$)
in DdDm~1, but were only able to set an upper limit on
the line flux. 
No lines were detected in H~4-1 (Fig. 2), and   
no continuum was detected for either object.  
Line fluxes are summarized in Table~1.

\subsection{Optical Lines}

For DdDm~1 we also utilized a high-dispersion, flux-calibrated
spectrum obtained on the McDonald Observatory 2.7~m Harlan
J. Smith Telescope, using the 2d-coud\'{e} 
white pupil cross-dispersed echelle spectrometer
(Tull et al. 1995). 
DdDm~1 was observed on 10 July 1995 for a
total integration time of 1 hour (60 min) at an air mass
of X = 1.01--1.04. 
Unfortunately, we did not observe H~4-1, so parameters
determined from optical lines are taken from the literature
for this object.

The data were reduced using the echelle package 
in IRAF.\footnote{The Image Reduction and Analysis Facilities
package is distributed by NOAO, which is operated by AURA, Inc.,
under contract to the National Science Foundation.}
The spectrum was extracted
over an effective area of 
1.8$\arcsec$ (slit width) $\times$ 
8$\arcsec$ (length). 
A rough wavelength calibration was established by observing
a ThAr lamp, although this is poorly constrained beyond
9000 \AA\, where we did not have accurate wavelengths for
the calibration lamp lines. There was no ambiguity, however,
in identifying the intrinsically strong lines of interest.

The spectral resolving power was R~$\sim$~60,000
(5~km~sec$^{-1}$). At this resolution, the nebular lines
are clearly resolved. The ionic lines such as [\ion{O}{3}]
4959 \AA\ and 4363 \AA\ have double-peaked profiles, with
a shallow (10--15\%) dip between two peaks of roughly equal
strength separated by 20.5 $\pm$ 0.3 km~s$^{-1}$. The FWHM
as measured from the half-power points on the extreme blue and
red sides of the overall profile is 41.0 $\pm$ 0.3 km~s$^{-1}$. 
Given the lower S/N for the infrared lines, their profiles
(Fig. 1) appear consistent with those of the optical lines.

Standard stars $\sigma$ Sgr (HR~7121)
and $\lambda$ Aql (HR 7236) were used with fluxes from 
Alekseeva et al. (1996) 
to construct the response function for flux calibration.
Due to atmospheric absorption
at the long-wavelength end ($\lambda$ $\ge$ 7000 \AA)
and crowding of Balmer lines in the calibration stars at
the short-wavelength end ($\lambda$ $\le$ 4000 \AA), as
well as the difficulty of accurately compensating for
differential refraction over such a 
broad wavelength range, the flux ratios of lines at
widely separated wavelengths are relatively uncertain.
However, flux ratios of lines in the same order, or
in adjacent orders for which the flux calibration can be
checked or scaled in the overlap region, should be
accurate to $\sim$10\%\ or better, which is 
sufficient for our purposes. In practice, we have 
used known intrinsic ratios of various 
\ion{H}{1} lines to adjust the fluxes of other
lines, and thereby infer additional line ratios (see \S 3.1.2).

The [\ion{S}{3}] 9069 \AA\ line, which we use below to
derive the ionic abundance ratio N(S$^{++}$)/N(H$^+$), 
deserves further comment. A slight asymmetry in its
profile as compared to the profiles of other ionic
lines (see above) indicates
a modest amount of absorption  
by telluric H$_2$O; 
the red side of the double-peaked profile is 
diminished by about 20\%\ relative to the blue side, and
the red half of the line profile is slightly narrower than
the blue side.  By comparing the blue half of the line profile
with the red half, we estimate that the line flux has been 
decreased by 10\%; we apply this correction to yield 
the [\ion{S}{3}] 9069 \AA\ flux reported below.

\section{RESULTS}

\subsection{DdDm~1}

\subsubsection{Nebular Parameters}

In order to compute ionic abundance ratios relative to hydrogen,
we need a measure of the H$^{+}$ emission. 
Since the apertures used for the observations were large enough to  
include the entire nebula and the extinction is negligible,
we use the total H$\beta$ flux for this comparison.
From our optical spectrum of DdDm~1 we find 
F(H$\beta$) = (1.52$\pm$0.03) $\times$ 10$^{-12}$
erg~cm$^{-2}$~s$^{-1}$, in excellent agreement with
B84 and Clegg et al. (1987; hereafter C87),
who give (1.51$\pm$0.20) $\times$ 10$^{-12}$ and
(1.45$\pm$0.15) $\times$ 10$^{-12}$
erg~cm$^{-2}$~s$^{-1}$ respectively. The extinction
towards this object is effectively zero (C87),
so we make no extinction corrections to the line fluxes.

We also need values for the nebular electron
temperature and density in order to compute effective
line emissivities. We adopted $T_e$ = 12,000 K and
$n_e$ = 4 $\times$ 10$^{3}$ cm$^{-3}$, after BC84 and C87.
The density-sensitive flux ratio of the [\ion{S}{2}]
lines is measurable in our optical spectra,
with both lines appearing in the same order, and 
we find F(6731 \AA)/F(6716 \AA) = 1.63$\pm$0.08,
consistent with the value of 1.7 found by these authors.

The lines which constitute the temperature-sensitive 
[\ion{O}{3}] ratio
F(4959 \AA)/F(4363 \AA) are 
separated by 10 echelle orders in our spectrum, so
it is not advisable to take their ratio directly. 
The intrinsically weaker [\ion{O}{3}] 4363 \AA\ 
line lies in the
same order as H$\gamma$, and our measured ratio
of these lines, 0.084$\pm$0.004 (see Table 2 for a summary
of the relevant optical line flux ratios),
is slightly lower than the values of 0.11 and 0.10
found by B84 and C87 respectively.  However, we also
find a slightly lower ratio of F([\ion{O}{3}]~4959 \AA)/H$\beta$
(1.10$\pm$0.05) than these authors (1.4--1.5 respectively),
so that after scaling by the intrinsic emissivity ratio of the 
H$\gamma$/H$\beta$ (0.47; Osterbrock 1989), 
we derive F(4959 \AA)/F(4363 \AA) $\sim$ 28$\pm$3,
whereas C87 and B83 find values of 32--33. 
Thus, our results are reasonably consistent with previous
temperature determinations.

\subsubsection{Sulfur}

The calculated value for the effective collision strength
of the [S IV] fine-structure line at 10.5 $\micron$ has 
increased dramatically over the past couple of decades,
altering the interpretation of earlier work on S$^{+3}$
in PNe (e.g. Dinerstein 1980a, 1980b; Beck et al. 1981).
At the present time, we have a more satisfactory situation, 
with the recent publication of two independent calculations 
which are in good agreement. 
The IRON-project calculations of Saraph \& Storey (1999) yield
$\Upsilon$(12,000 $K$) = 8.55. The value reported by
Tayal (2000) for $T$ = 10,000 K,   
$\Upsilon$ = 8.536, is consistent with that of Saraph \& Storey.

Using this collision strength for the nebular parameters cited 
above, the density-normalized volume emissivity for the [\ion{S}{4}] line is
(Ah$\nu$f$_2$)/$n_e$ = 4.6 $\times$ 10$^{-20}$ erg~cm$^{3}$~s$^{-1}$,
where A is the transition probability, $\nu$ is the 
frequency of the emitted line,  f$_2$ is the 
fraction of ions in the upper level of the transition, and
the other symbols have their usual meanings.
From our line measurement, this yields 
S$^{+3}$/H$^+$ = (2.7$\pm$0.2) $\times$ 10$^{-7}$, 
where the error bar includes only the uncertainties 
in the infrared line flux.

The central star of DdDm~1 is relatively cool, similar 
to that of the M15 PN K648. Pe\~{n}a et al. (1992) find
$T_{*}$(DdDm~1) = 40,000$\pm$5000 K 
and $T_{*}$(K648) = 35,000$\pm$5000 K 
from UV and optical measurements of the stellar continuum, while
Howard et al. (1997) calculated photoionization models based on
observations from the literature and found best-fit values of
$T_{*}$ = 45,000 K for both objects.
At these low stellar temperatures, one expects that most
of the sulfur would be doubly, not triply ionized.

The [\ion{S}{3}] lines at 9069 and 9531 \AA\ are among   
the strongest lines in the echelle spectrum of DdDm~1.
Fortuitously, each line is paired
with a recombination line of \ion{H}{1} 
that appears in the same order: [\ion{S}{3}] 9069 \AA\
with 9015 \AA\ (Pa~10) and [\ion{S}{3}] 9531 \AA\ with 
9229 \AA\ (Pa~9). However, this region of the spectrum is
severely affected by telluric H$_2$O absorption, and
the wavelength position of a particular line --
which depends on both the object's intrinsic radial velocity shift
and the date-dependent geocentric correction -- becomes
critical. At our high resolving power, R $\sim$ 60,000,
we can clearly see telluric absorption in the spectra of our
calibration stars, and are therefore able to assess its impact
on the line measurements. For the date of
the DdDm~1 observations, absorption by H$_2$O had
only a minor effect on the measured flux of 
[\ion{S}{3}] 9069 \AA\ (see \S 2.2 above), but  
[\ion{S}{3}] 9531 \AA\ is severely affected, with
part of the line profile falling in a region of
essentially zero transmission by the atmosphere.

We therefore use the measured ratio
of F([\ion{S}{3}] 9069 \AA)/F(\ion{H}{1} Pa~10) = 
9.5$\pm$0.09, together with the [\ion{S}{3}]
and H$\beta$ emissivities from the web-based multi-atom
program of Shaw \& Dufour (1995) and the emissitivity ratio 
of Pa~10/H$\beta$ from Osterbrock (1989), and derive
S$^{++}$/H$^+$ = (1.63$\pm$0.16) $\times$ 10$^{-6}$.
Previously published values for S$^{++}$/H$^+$ were based 
on the weaker, more temperature-sensitive
[\ion{S}{3}] 6312 \AA\ line: 
(2.3$\pm$0.5) $\times$ 10$^{-6}$ (B84); and 
2.56 $\times$ 10$^{-6}$ (C87). Unfortunately, 
this line is not present
in our data because it falls in a gap between echelle orders.
However, all of these values are consistent with the
expectation that S$^{++}$ is the dominant ionization stage
of S in DdDm~1. Consequently, the uncertainty in the
total S/H abundance is dominated by the uncertainty in
S$^{++}$/H$^+$. Not explicitly included in the cited
errors is any systematic uncertainty due to 
the choice of atomic parameters A and $\Upsilon$
(we used values from Mendoza \& Zeippen 1982 and 
Galavis, Mendoza, \& Zeippen 1995).

We derived the abundance of S$^{+}$ 
from the ratio F([\ion{S}{2}] 6731 \AA)/F(H$\beta$) 
(Table 2) and the multi-level atom predictions
of Shaw \& Dufour (1995), finding 
S$^{+}$/H$^+$ = (2.2$\pm$0.2) $\times$ 10$^{-7}$.
(We were unable to use H$\alpha$, which lies closer
in wavelength, because it fell too close to the edge
of the echelle order.)
This value is intermediate
between those found by B84 (1.6 $\times$ 10$^{-7}$)
and C87 (2.7 $\times$ 10$^{-7}$)
and comparable to
the amount of S$^{+3}$/H$^+$. Thus, the data bear out
our expectation that S$^{++}$ is the dominant ion of S in DdDm~1.

The total S/H abundance in DdDm~1 summed over the three main
ionization states (Table 3) is thus (2.1$\pm$0.2) $\times$ 10$^{-6}$. 
This is about an order of
magnitude lower than S/H in typical disk PNs 
but an order of magnitude higher than for the extreme
halo PNs BB~1, K648, and H~4-1 as measured from
S$^+$ and S$^{++}$ (Barker 1983; also see discussion below).
The S/O ratio 
is less sensitive than S/H to the adopted 
nebular parameters because of the similar temperature
dependence of the line emissitivies. Depending on whether
we adopt O/H = 1.1 $\times$ 10$^{-4}$ (B84) or
O/H = 1.4 $\times$ 10$^{-4}$ (C87), we have 
S/O = 0.015--0.019. 
This is larger than values for disk PNe found
by Kwitter \& Henry (2001) and Milingo, Henry, \& Kwitter
(2002), who give S/O = .012$\pm$.007 and .011$\pm$.006
respectively, 
but did not measure S$^{+3}$. 
On the other hand, our S/O is only  
slightly below the value of .021,
determined from ISO measurements of fine-structure 
lines of both [\ion{S}{3}] and [\ion{S}{4}],
for a metal-poor extragalatic \ion{H}{2} region
which has similar O/H to that DdDm~1 
(Nollenberg et al. 2002).

\subsubsection{Neon}

>From the ratio of our [\ion{Ne}{2}] 12.8 $\micron$ flux 
to H$\beta$, using a collision
strength of $\Upsilon$ = 0.306  (interpolated for $T_e$ = 12,000 K, based
on Johnson \& Kingston 1987; see Pradhan \& Peng 1995), we derive
Ne$^{+}$/H$^{+}$ = (2.13$\pm$0.2) $\times$ 10$^{-6}$. 
We find Ne$^{++}$/H$^{+}$ = (1.17$\pm$0.04) $\times$ 10$^{-5}$. 
from F([\ion{Ne}{3}] 3967 \AA)/F(H$\epsilon$) = 0.55$\pm$.02,
which are separated by only 3 \AA\ and hence are in the
same echelle order, and H$\epsilon$/H$\beta$ = 0.16
(Osterbrock 1989). We also measured [\ion{Ne}{3}] 3869 \AA,
which lies two orders away from 3967 \AA, and find 
F(3869 \AA)/F(3967 \AA) = 2.52$\pm$0.07, in reasonable
agreement with the theoretical ratio of 2.4 (set by
the A values since both lines arise from the same level),
given the uncertainties in comparing lines in different echelle orders.
(The 3869 \AA\ line is in the same order as H8, but 
a cosmic ray hit interferes with our measuring H8.)

Our value of (1.17$\pm$0.04) $\times$ 10$^{-5}$ 
can be compared with the results of 
B84, who found Ne$^{++}$/H$^{+}$ = (1.5$\pm$0.3) $\times$ 10$^{-5}$
and C87, who derived 1.3 $\times$ 10$^{-5}$. Both of these groups,
however, applied a fairly large ionization correction factor 
for the unobserved Ne$^+$, which raised their elemental 
abundance ratio to Ne/H = 2.0 $\times$ 10$^{-5}$. 
Since we find Ne$^{+}$/Ne$^{++}$ = 0.18$\pm$.04,
our value for the elemenetal abundance,
Ne/H = (1.4$\pm$0.1) $\times$ 10$^{-5}$,  
is 25\%\ lower than theirs.
Corrections for Ne$^{+}$ based on photoionization models
apparently remain problematical (e.g. Masegosa et al. 1994).   

We consider the contribution of Ne$^{+3}$ to be negligible,
in view of the high ionization potential of Ne$^{++}$ (63.45 eV)
and complete absence of \ion{He}{2} in the spectrum of DdDm~1
(F(\ion{He}{2} 4686 \AA)/F(H$\beta$) $<$ 1.1 $\times$ 10$^{-3}$,
3$\sigma$). Likewise, the neutral ion, Ne$^0$ is unlikely to be important,
given the relatively small amount of Ne$^{+}$. The total Ne/H 
abundance is then Ne/H = 1.4$\pm$0.1 $\times$ 10$^{-5}$, and
Ne/O = 0.10--0.13, depending on the adopted value of O/H. 

\subsubsection{Limit on Hu~$\alpha$}

We also attempted to measure
an \ion{H}{1} recombination line in the mid-IR, in order
to provide a comparison line observed with the same instrument.
The strongest such line available in the mid-infrared is
\ion{H}{1} 7--6 (Hu~$\alpha$), at 12.37 $\micron$. In our limited
observing time, we set a 3$\sigma$ upper limit of 
$\sim$ 3 $\times$ 10$^{-14}$ erg~cm$^{-2}$~s$^{-1}$ on the line
flux. However, this is consistent with the expected line strength.
For Case B recombination at $T_e$ = 10,000 K, $n_e$ = 10$^4$ cm$^{-3}$,
the ratio F(Hu~$\alpha$)/F(H$\beta$) = 9.3 $\times$ 10$^{-2}$
(Hummer \& Storey 1987),
from which we predict F(Hu~$\alpha$) $\sim$ 1.4 $\times$ 10$^{-14}$,
a factor of two smaller than our limit. This upper limit does
preclude, however, any unexpectedly large extinction towards
the source.

\subsection{H~4-1}

H~4-1, sometimes designated by its galactic coordinates as 49+88,
was one of the first two Type IV PNe discovered. The other,
the PN commonly designated either as K648 or Ps 1 (Pease 1927),
is located in the metal-poor globular cluster M 15.  
Along with BB~1 (also known as BoBn~1 and 108-76), these three objects
were the focus of many optical studies 
(e.g. Torres-Peimbert \& Peimbert 1979; Barker 1980, 1983), prior to 
the discovery of several additional metal-poor PNe (e.g. Pe\~{n}a
et al. 1989, 1991, 1992).

The H$\beta$ flux of H~4-1 was measured by Hawley \& Miller (1978) 
as F(H$\beta$) = 2.63 $\times$ 10$^{-13}$ 
in a 2.4$\arcsec$ $\times$ 4.0$\arcsec$ aperture,  
which is comparable to our IR aperture. 
They find that the extinction is low, as expected
for an object located almost at the north Galactic pole; so,
as for DdDm~1, we ignore extinction corrections.   
The physical parameters for H~4-1 reported by Hawley \& Miller (1978)
and Torres-Peimbert \& Peimbert (1979), are similar to those of
DdDm~1, except for a somewhat lower density. We therefore
adopt $T_e$ = 12,000 K, $n_e$ = 1 $\times$ 10$^{3}$ cm$^{-3}$.

Our upper limit on [\ion{S}{4}] 10.5 $\micron$ in H~4-1 is 
5 $\times$ 10$^{-14}$ erg~cm$^{-2}$~s$^{-1}$ (3~$\sigma$). This
corresponds to S$^{+3}$/H$^{+}$ $\le$ 
3.8 $\times$ 10$^{-7}$. The abundance of
S$^{++}$ was reported as S$^{++}$/H$^{+}$ = 
(1.1$\pm$0.7) $\times$ 10$^{-7}$
by Barker (1983; hereafter B83), 
from a measurement of [\ion{S}{3}] 9531 \AA.
Torres-Peimbert \& Peimbert (1979) 
measured only S$^{+}$, which contains only a small fraction of
the total S in this PN, S$^{+}$/H$^{+}$ = 2.5 $\times$ 10$^{-8}$,
consistent with Barker's value of (2.7$\pm$0.8) $\times$ 10$^{-8}$.
Adding up the ionic ratios, we find an upper
limit of S/H = (S$^{+}$+S$^{++}$+S$^{+3}$)/H $\le$ 
5 $\times$ 10$^{-7}$. 
Taking O/H = 1.2 $\times$ 10$^{-4}$ (Howard et al. 1987), 
this corresponds to S/O $\le$ 4 $\times$ 10$^{-3}$. 
(Although Howard et al. derive S/H = 1.1 $\times$ 10$^{-7}$,
this is constrained only by the [\ion{S}{2}] lines, and
is likely to be quite uncertain for such a highly ionized nebula.)
This is comparable to the values of
S/O $\le$ 2 $\times$ 10$^{-3}$ and $\le$ 6 $\times$ 10$^{-3}$ 
found for K648 and BB~1 respectively by  
Garnett \& Lacy (1993), who also searched
for [\ion{S}{4}] IR line emission.

For [\ion{Ne}{2}] 12.8 $\micron$, our 3~$\sigma$ upper limit
of 4 $\times$ 10$^{-14}$ erg~cm$^{-2}$~s$^{-1}$ translates to
Ne$^{+}$/H$^{+}$ $\le$  5 $\times$ 10$^{-6}$. Singly-ionized Ne
is not expected to be the dominant ion, so it is not surprising
that we do not detect [\ion{Ne}{2}].  The abundance of
Ne$^{++}$ was found to be 
Ne$^{++}$/H$^{+}$ = 3 $\times$ 10$^{-6}$ by both 
Hawley \& Miller (1978) and Torres-Peimbert \& Peimbert (1979).
Our upper limit rules out the possibility of the elemental 
abundance being more than a factor of 2.6 larger, 
leaving previous conclusions about elemental abundance ratios
in H~4-1 essentially unchanged.

\section{DISCUSSION}

Metal-poor Galactic planetary nebulae are sufficiently rare that
each discovery of a new example is regarded as noteworthy. 
Currently there are only $\sim$ 10--12 objects considered to belong
to this class (Pe\~{n}a et al. 1992; Howard et al. 1997), with 
3--4 of these objects located in globular clusters (Pease 1928;
Gillett et al. 1989; Cohen \& Gillett 1989; Jacoby et al. 1997).
The deficit of PNe in globular 
clusters has long been regarded as puzzling 
(R. Kraft 1976, private communication).
It may be a consequence of the details of the mass-loss process
which creates an observable planetary nebula, which is not well 
understood. For example, low-metallicity stars may only produce
PNe with low masses; the ionized masses of the PNe in M~15 and
M~22 are $\sim$ 0.02 M$_\sun$ (see references above). Another 
possibility is that the Galactic globular clusters are too
old, and hence their turn-off masses 
too low, to produce a visible PN via normal, single-star evolution, 
and a more exotic evolutionary channel such as binary interactions
or coalescence is required to produce a PN (Jacoby et al. 1997, 1998;
Alves, Bond, \& Livio 2000).

A difficulty in unambiguously identifying a candidate object 
as a bona fide halo population PN is that the initial metallicity
determination is usually based on O/H, 
since O produces some of the strongest emission lines in their spectra.
There is an overall discrepancy or ``offset'' between most determinations
of nebular O/H and the solar value, such that nebular analyses 
adopting the standard ``isothermal'' assumption yield
derived O/H abundances of up to a factor of 2--3 lower than solar
(e.g. Dinerstein 1990; Peimbert 1995). 
Recent studies indicating a lower solar abundance,
O/H $\sim$ 5 $\times$ 10$^{-5}$  (Allende Prieto, Lambert, \&
Asplund 2001) diminish but do not entirely remove this discrepancy.  
However, it has been well-established that the ratios of
elements such as C, Ar, and S to O vary dramatically among the
halo population PNe (see references given in \S 1).
Furthermore, the fact that the PN in M~15, has 
O/H = 5 $\times$ 10$^{-5}$, which corresponds to 
[O/H] = --1.2 in the usual ``bracket'' notation where [O/H] =
log (O/H) -- log (O/H)$_{\sun}$, while the stars in 
the parent cluster have [Fe/H] = --2.2 (Harris \& Harris 2000),
reinforces the point that O/H may not be a reliable
index of general metallicity. 
This could be due either to internal enrichment
or to a more rapid
build-up of O as compared to the heavier $\alpha$-capture and
Fe-group nuclei over Galactic history (Barker 1980; Torres-Peimbert
et al. 1981, and references therein).

Low-metallicity PNe can also be found within the metal-poor
stellar populations of Local Group galaxies such as the LMC
and SMC (Dopita et al. 1997; Jacoby \& De Marco 2002). For
the LMC sample, the heavier $\alpha$-elements appear to be
reliable indicators of metallicity. In particular, Ne does
not appear to be affected by internal processing and dredge-up. 
More recently, P\'{e}quignot et al. (2000a, b) found evidence 
of oxygen self-enrichment in two PNe belonging to 
the Sagittarius dwarf spheroidal galaxy which is
merging with the Milky Way, based on low values of  
Ne/O, S/O, and Ar/O which appear to result from elevated O/H.
This agrees with the current picture that dredge-up processes
are more efficient in metal-poor stars than
in those of near-solar metallicity. 

Tovmassian et al. (2001) reported the discovery that the
previously catalogued emission-line object SBS 1150+599A is 
an extremely oxygen-poor PN, now designated PNG 135.9+55.9. 
This unusual object shares many of the characteristics of
the high-ionization halo PNe NGC 2242 and NGC 4361, in which
the \ion{He}{2} lines are strong and \ion{He}{1} is absent. 
The weakness of the [\ion{O}{3}] 4959, 5007 \AA\ lines in 
NGC 2242 and 4361  is due primarily to ionization effects, and  
these objects have been interpreted as optically thin nebulae 
which are mass-bounded with the He$^{++}$ zone (Garnett \& Dinerstein
1988; Torres-Peimbert, Peimbert, \& Pe\~{n}a 1990; Liu 1998).
Follow-up studies of PNG 135.9+55.9 by Richer et al. (2002)
and Jacoby et al. (2002) appear to confirm that the O/H
abundance is low, $\le$ .01 solar, although the exact value
is somewhat model-dependent. The abundances of other elements
are harder to establish: Richer et al. find Ne/O = 0.5$\pm$0.3,
while Jacoby et al. derive Ne/O = 2.3--4.3. The heavier
$\alpha$ elements Ar and S are still more difficult, since
in the optical they are represented only by lines that are
weak or from minority ions. 
Another possibility is that this object
has experienced a separation of gas and dust leading to
deficiencies in selected elements, an effect seen in some
post-AGB stars (e.g. Napiwotzki, Heber, \& K\"{o}ppen 1994).
Clearly, PNG 135.9+55.9 deserves further attention.

The initial sample of Galactic Type IV PNe, K648, H~4-1, BB~1, and
DdDm~1, all have O/H within a factor of two of $\sim$ 10$^{-4}$,
and the additional halo PNe discovered subsequently also
have O/H in this range. However, as discussed above, they
show a wide dispersion in the ratios of heavier $\alpha$-capture elements to O.
From optical studies, BB~1, K648, and H~4-1 
had been found to have very low values of S/O, 
while DdDm 1 showed an S/O value similar to that of disk PNe. 
In view of the fact that the correction  
for S$^{+3}$ from ionization models was uncertain,
Garnett \&  Lacy (1993) addressed this issue empirically by
searching for [\ion{S}{4}] 10.5 $\micron$ in 
K648 and BB~1. They failed to detect the line in either object,
setting upper limits of S$^{+3}$/H$^{+}$ $\le$ 1 $\times$ 10$^{-7}$ and
3 $\times$ 10$^{-7}$ for K648 and BB~1 respectively, and
corresponding limits of 
S/O $\le$ 2 $\times$ 10$^{-3}$ and $\le$ 6 $\times$ 10$^{-3}$.
Garnett \& Lacy interpreted these results in terms of 
elevated O rather than anomalously low S, and suggested that
the enhancement
in O/H is due to nucleosynthesis and mixing in the progenitor star.

Our observations supplement and extend those of Garnett \& Lacy
by including two additional halo PNe: H~4-1, which appears to belong
to the same extremely metal-poor population as BB~1 and K648;  
and DdDm~1, which is characterized by relative abundances of 
S/O and Ar/O similar to those of ordinary disk PNe.   
We have also observed the ``missing'' ion of Ne, which is
a lower ionization potential stage (Ne$^+$) in addition to
the missing higher ionization stage of S (S$^{+3}$). 
The two objects in the present study are well matched
in excitation and stellar temperatures to the two PNe observed
by Garnett \& Lacy, with DdDm~1 being similar to K648 and
H~4-1 having similar excitation as BB~1 (see \S 3.1.2). 
With this more complete set of observations, the conclusion
that there are two distinct
types of metal-poor Galactic PNe seems inescapable.

The planetary nebula populations of external galaxies are being
tapped as valuable tracers of a number of chacteristics of their
parent galaxies: as distance indicators through match-up of
their luminosity functions (e.g. Jacoby et al. 1992); as kinematical tracers
of the gravitational potential wells of the spheroids
of external galaxies (e.g. Napolitano et al. 2001); and even
as metallicity indicators (Walsh et al. 2000).
However, it is becoming clear that O/H 
abundances in the PN gas are more strongly affected by self-enrichment
than previously suspected. Indeed, these processes may effectively
set an abundance ``floor'' for O/H which erases any original 
one-to-one correlation between O/H and the abundances of heavier
elements such as Ar, S, and Fe.  This erasure is presumably 
responsible for the wide scatter in ratios such as Ar/O and S/O
among the halo PNe (Pe\~{n}a et al. 1992; Howard et al. 1997).
The moral to be drawn is that caution
must be applied when using O/H as a generic abundance indicator
for PNe, particularly in low-metallicity parent
populations; it is of utmost importance than abundances be
measured for elements heavier than the CNO group, in order to
establish the original composition of the their stellar progenitors.

\section{CONCLUSIONS}

We have observed the fine-structure lines [\ion{S}{4}] 10.5 $\micron$
and [\ion{Ne}{2}] 12.8 $\micron$ in two PNe with similarly 
low O/H abundances, O/H $\sim$ 10$^{-4}$; these objects
belong to the small sample of Galactic PNe comprising 
the class of halo population or Type IV nebulae.
We detect both lines in DdDm~1, and set upper 
limits on their fluxes in PN H~4-1.
These observations, combined with new (for DdDm~1) and
previous (H~4-1) optical data, confirm that these two objects
have distinctly different patterns of heavy element abundances,
despite their similar values of 
O/H, which has traditionally been regarded as a reliable indicator of
metallicity in ionized nebulae. As others have
previously suggested, it appears that H~4-1, along with the
similar objects K648 and BB~1 observed by Garnett \& Lacy (1993),
has self-enriched its nebular gas with oxygen synthesized
within the star. 
In these objects, it is the heavier elements
such as S and Ar which more accurately
reflect the initial stellar composition. In contrast, 
DdDm~1 has similar heavy-element ratios to those of disk PNe, and
represents a distinct, only modestly metal-poor population, 
to which several
other recently discovered Galactic halo PNe (e.g. M~2-29, 
PRGM~1) also belong.

These results suggest that O/H should be approached
with caution as an indicator of metallicity in
metal-poor PNe. The study of 
PNe in external galaxies has been rapidly expanding,
from their initial use as distance indicators, to
kinematical tracers of galaxy potential wells 
and indicators of the composition of their parent populations.
However, in view of the discussion above, it is possible 
that extremely low O/H values may be very rare even in PNe
from very metal-poor populations, because of 
efficient dredge-up and consequent self-enrichment. 
Further studies of metal-poor PNe, both 
in the Galaxy and in external galaxies, will be very
useful for clarifying this issue and for helping
determine the extent to which lower-mass stars are
significant sources of oxygen to the ISM.

We thank the day and night staff of the IRTF for their help with
the TEXES observations, and G. Jacoby for
helpful discussions.  
J. Crawford, C. Sneden, R. French, K. Volk, 
and E. Brugamyer assisted in obtaining and reducing
the optical spectrum of DdDm~1; R. Shaw, A. Pradhan, and
N. Sterling helped provide line emissivities.
HLD was supported by NSF grant AST 97-31156 and a Faculty
Research Assignment (sabbatical) grant. Support for MJR and for
development of TEXES was provided by grants from the NSF,
NASA through USRA, and 
the Texas Advanced Research Program (TARP).

\begin{deluxetable}{llcccc}
\tablecolumns{5}
\tablewidth{0pc} 
\tablenum{1}
\tablecaption{Infrared Line Fluxes}
\tablehead{
\colhead{Nebula} & \colhead{Ion/Line} & \colhead{Flux} & \colhead{Date of} & \colhead{Exp. Time} \\
\colhead{}  & \colhead{} & \colhead{(erg~cm$^{-2}$~s$^{-1}$)} & \colhead{Obs.} & \colhead{(sec)}}
\startdata
DdDm~1... & [\ion{Ne}{2}] 12.8 $\micron$ & 1.0$\pm$0.1 $\times$ 10$^{-13}$ & 29 June 2001 &  580\\
	& [\ion{S}{4}] 10.5 $\micron$ & 1.75$\pm$.08 $\times$ 10$^{-13}$ & 30 June 2001  &  580\\
	& \ion{H}{1}~7--6 12.4 $\micron$ & $<$ 3 $\times$ 10$^{-14}$ & 29 June 2001 &  290\\
H~4-1...... & [\ion{Ne}{2}] 12.8 $\micron$ & $<$ 4 $\times$ 10$^{-14}$ & 30 June 2001 &  200\\
	& [\ion{S}{4}] 10.5 $\micron$ & $<$ 5 $\times$ 10$^{-14}$ & 29 June 2001 &  375\\
\enddata
\end{deluxetable}

\clearpage 
 
\begin{deluxetable}{lc}
\tablecolumns{2}
\tablewidth{0pc}
\tablenum{2}
\tablecaption{Optical Line Ratios in DdDm~1\tablenotemark{a}}
\tablehead{
\colhead{Transitions} & \colhead{Ratio}}
\startdata
F([\ion{O}{3}]~4363)/F(H$\gamma$)  			&  0.084$\pm$.004   \\
F([\ion{O}{3}]~4959)/F(H$\beta$)\tablenotemark{b}  	&  1.10$\pm$0.05   \\
F([\ion{S}{2}]~6731)/F([\ion{S}{2}]~6716) 		&  1.63$\pm$0.08  \\
F([\ion{S}{2}]~6731)/F(H$\beta$)\tablenotemark{c}	&  0.049$\pm$.003  \\
F([\ion{S}{3}]~9069)/F(Pa~10)   			&  9.5$\pm$0.9   \\
F([\ion{Ne}{3}]~3967)/F(H$\epsilon$)  			&  0.55$\pm$0.02   \\
F([\ion{Ne}{3}]~3869)/F([\ion{Ne}{3}]~3967)\tablenotemark{b} &  2.52$\pm$0.07   \\
\tablenotetext{a}{Unless otherwise noted, the lines are in the same echelle order.}
\tablenotetext{b}{Separated by 2 echelle orders.}
\tablenotetext{c}{Separated by 20 echelle orders.}
\enddata
\end{deluxetable}

\clearpage

\begin{deluxetable}{lcccc}
\tablecolumns{3} 
\tablewidth{0pc} 
\tablenum{3}
\tablecaption{Ionic and Total Abundances}
\tablehead{
\colhead{Ratio} & \colhead{DdDm~1} & \colhead{H~4-1}}
\startdata
S$^{+}$/H$^+$  & 2.2$\pm$0.3 $\times$ 10$^{-7}$  & 2.5 $\times$ 10$^{-8}$  \\
S$^{++}$/H$^+$  & 1.63$\pm$0.16 $\times$ 10$^{-6}$ & 1.1 $\times$ 10$^{-7}$  \\
S$^{+3}$/H$^+$  & 2.3$\pm$0.2 $\times$ 10$^{-7}$ & $\le$ 3.8 $\times$ 10$^{-7}$  \\
Total S/H....   & 2.1$\pm$0.2 $\times$ 10$^{-6}$   & $\le$ 5 $\times$ 10$^{-7}$ 	\\
Ne$^{+}$/H$^+$  & 2.1$\pm$0.2 $\times$ 10$^{-6}$  & $\le$ 4.7 $\times$ 10$^{-6}$  \\
Ne$^{++}$/H$^+$  & 1.17$\pm$0.04 $\times$ 10$^{-5}$  &  3.0 $\times$ 10$^{-6}$  \\
Total Ne/H....   & 1.4$\pm$0.1 $\times$ 10$^{-5}$   & $\le$ 8 $\times$ 10$^{-6}$ \\
\enddata 
\end{deluxetable}

\clearpage

\figcaption[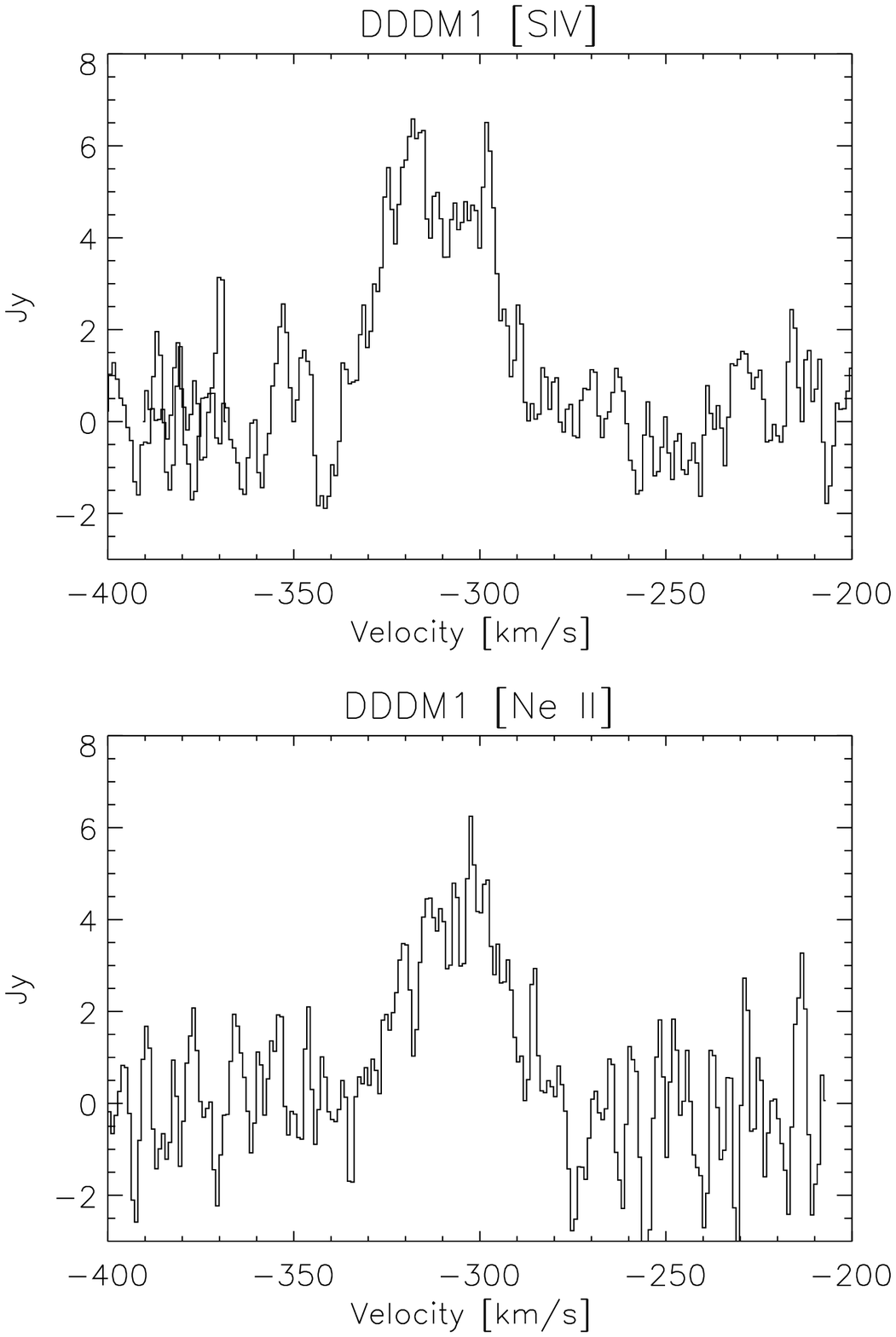]{Spectra of DdDm~1 obtained with TEXES on the IRTF
in the regions of [\ion{S}{4}] 10.5 $\micron$ (top) and [\ion{Ne}{2}] 12.8 $\micron$ (bottom).
Both lines appear at the expected wavelengths given the nebula's radial velocity
of --304 km~s$^{-1}$. The lines are 
spectrally resolved, with FWHM values of $\sim$ 30--40 km~s$^{-1}$.
The brighter line, [\ion{S}{4}], shows the double-peaked structure typical
of ionic lines in this PN (see \S 2.2); [\ion{Ne}{2}] is noisier and
the shape of its profile is less certain.} 
 
\figcaption[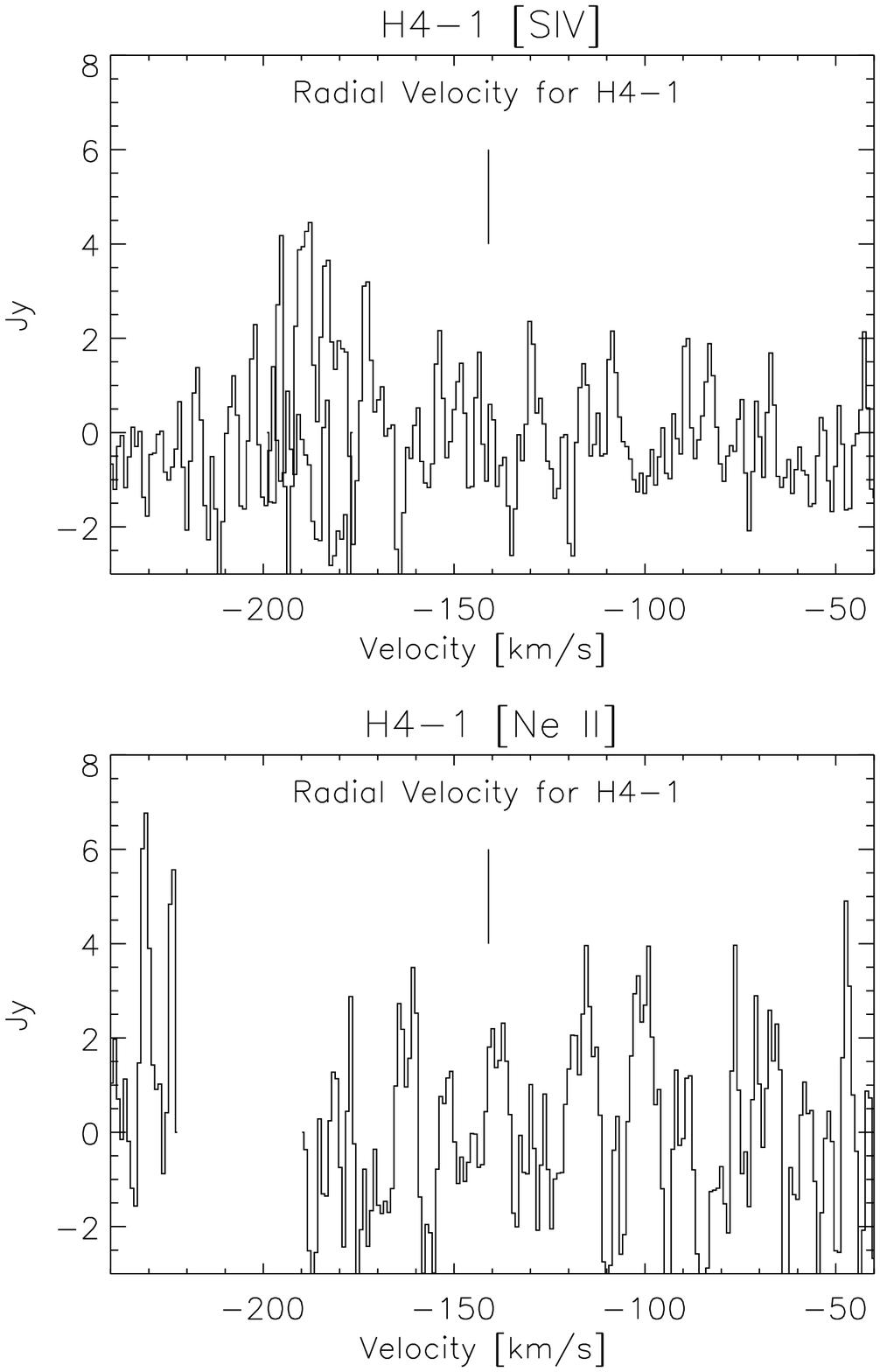]{The regions of the 
[\ion{S}{4}] 10.5 $\micron$ (top) and [\ion{Ne}{2}] 12.8 $\micron$ (bottom) 
are shown for H~4-1, in which the lines were not detected.
The expected wavelengths for this nebula's radial velocity
of --141 km~s$^{-1}$ are indicated by vertical tick marks. 
Two orders overlap in the region of the [\ion{S}{4}] line, and
there is a gap near [\ion{Ne}{2}], but neither would interfere
with detection of the lines.}

\end{document}